\documentclass[3p,twocolumn]{elsarticle}

\usepackage{graphicx}
\usepackage{dcolumn}
\usepackage{bm}
\usepackage{amsmath}
\usepackage{amssymb,amsthm}
\usepackage{color}
\usepackage{epstopdf} 
\usepackage[sans]{dsfont}
\usepackage{accents}
\usepackage[utf8]{inputenc}
\usepackage[T1]{fontenc}
\usepackage[english]{babel}
\usepackage{hyperref}
\usepackage{mathbbol,bbm}
\usepackage{float}
\usepackage{accents}
\usepackage{mathtools} 
\usepackage{booktabs}

\def\ket#1{\mathinner{|{#1}\rangle}}
\newcommand{\braket}[2]{\langle #1|#2\rangle}
\newcommand{\ketbra}[2]{\left|#1\rangle\langle #2\right|}

\newcommand{\matelem}[3]{\langle #1|#2|#3\rangle}

\newcommand{\ketbbra}[2]{\left|#1\rangle\langle\langle #2\right|}
\def\bbra#1{\langle\langle#1|}

\newcommand{\co}[1]{\cos\left(#1\right)}
\newcommand{\si}[1]{\sin\left(#1\right)}
\newcommand{\piN}[1]{\frac{#1 \pi}{N+1}}

\graphicspath{{FigsEfield/}}

\begin{document}

\title{Electric-field assisted optimal quantum transport of photo-excitations in polar heterostructures}


\author[1,2,3]{Chahan M. Kropf\corref{cor1}}
\ead{Chahan.Kropf@gmail.com}

\author[4]{Giuseppe Luca Celardo}

\author[2,3]{Claudio Giannetti} 

\author[1,2,3]{Fausto Borgonovi} 
\ead{Fausto.Borgonovi@unicatt.it}

\address[1]{Istituto Nazionale di Fisica Nucleare, Sezione di Pavia, via Bassi 6, I-27100 Pavia, Italy
}
\address[2]{Department of Physics, Universit\`a Cattolica del Sacro Cuore, Brescia I-25121, Italy}
\address[3]{ILAMP (Interdisciplinary Laboratories for Advanced Materials Physics), Universit\`a Cattolica del Sacro Cuore, Brescia I-25121, Italy} 
\address[4]{Benem\'erita Universidad Aut\'onoma de Puebla, Apartado Postal J-48, Instituto de F\'isica, 72570, Mexico}

\cortext[cor1]{Corresponding author}


\date{\today}

\begin{abstract}
Transition-metal-oxide (TMO) heterostructures are promising candidates for building photon-harvesting devices which can exploit optimal quantum transport of charge excitations generated by light absorption. Here we address the explicit role of an electric field on the quantum transport properties of photo-excitations subject to dephasing in one-dimensional chains coupled to a continuum of states acting as a sink. We show that the average transfer time to the sink is optimized for suitable values of both the coupling strength to the sink and the electric field, thus fully exploiting the coherence-enhanced efficiency in the quantum transport regime achievable in few monolayers TMO heterostructures. The optimal coupling to the continuum remains approximately the same as that in absence of electric field and is characterizing the Superradiant Transition. On the other hand, the optimal electric field for which we provide estimates using an analytical expression is dependent on the initial state.
\end{abstract}

\maketitle


\section{Introduction}

Transition-metal-oxides (TMOs) heterostructures have been used to build solar-cells \cite{Chang2016} and have been proposed as materials for achieving unprecedented light-conversion efficiency. As paradigmatic example of polar TMOs, LaVO$_3$ \cite{Assmann2013} exhibits a correlation-driven direct Mott gap in the optimal range for solar light absorption and an internal electric potential gradient that was argued to help separating the photo-generated electron-hole pairs. Recently it was discussed that with proper engineering of the collector properties, the system can be tuned to the superradiant transition which gives rise to fast quantum coherent transport in thin films made of few TMO monolayers \cite{Kropf2019}. This effect was shown to be robust against realistic estimates of electron-electron correlations \cite{Kropf2019} and to environment induced dephasing up to ambient temperatures. The effect of the intrinsic electric potential gradient was taken into account only insofar as a shift of the Fermi-level of the collector as is common in the Landauer-Büttiker \cite{Ryndyk2016,Landauer1957,Pastawski1991} and the master equation \cite{Gurvitz1996} approaches to transport. However, an intrinsic potential slope also leads to a constant shift of the energy levels. While in the classical diffusive-like regime such a shift has only little influence, this is no longer the case in the quantum coherent regime where one obtains for instance, Bloch-oscillations in the infinite size limit \cite{Zener1934}. Furthermore, one cannot directly transfer the intuition from the semi-classical regime that an electric field favors transport by accelerating the charge carriers towards the collector. Hence, we expect potentially large changes in the transport properties. Here we study the influence of the electric potential gradient inside the system on the transport properties of few-monolayers TMO heterostructures with a particular emphasis on the quantum-coherent regime. Our primary aim is to clarify whether, and for which values of the electric field the previously described superradiant quantum coherent transport (in absence of electric field, see \cite{Kropf2019}) prevails that gives rise to coherence-driven enhancement of the photo-transport efficiency. In addition, we seek to understand if the electric field can be tuned to further optimize photo-transport.

\begin{figure}[t]
	\centering
	\includegraphics[width = 0.5\textwidth]{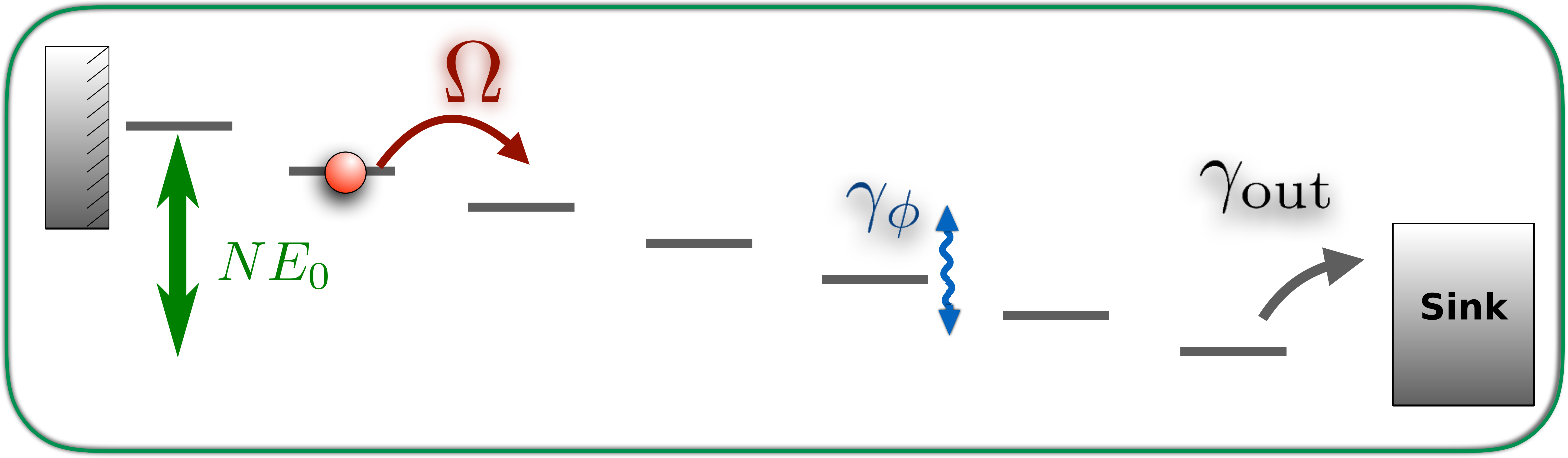}
	\caption{Transport model of a single excitation hopping through a chain with nearest-neighbour coupling $\Omega$ with a decay rate $\gamma_{out}$ to the sink. The chain is subject to a constant potential $N E_0$ and pure dephasing on each site with rate $\gamma_\phi$.}
	\label{fig:Fig1}
\end{figure}

As a realistic, though simplified, platform we investigate the average transfer time to the sink of an excitation in a one-dimensional chain with an intrinsic constant potential gradient and nearest-neighbour interactions as illustrated in Fig.~\ref{fig:Fig1}. The influence of further environmental degrees of freedom, such as phonons or dynamical noise, are modelled by the inclusion of an effective pure dephasing channel on each site. We show that for electric potentials smaller than the mean-level spacing the transport is quantum coherent and optimized at a coupling to the sink correspondent to the superradiant transition (ST). Hence, superradiance emerges as an extremely valid transport optimization principle \cite{Celardo2012}, which is robust against disorder \cite{Zhang2017a,Zhang2017b},  decay, dephasing, electron-electron interactions \cite{Kropf2019} and a moderate electric potential.

In our study we consider mainly  the average transfer time for an initial Gaussian wavepacket. The case of an initial state localized on a single site is also briefly discussed to mark the differences with the former case. We show that in the quantum coherent regime the electric field mainly affects transport through the modification of the shape of the eigenstates in the site basis. Indeed, for a given initial state, the electric field can be tuned to maximize (or minimize) the overlap with the faster decaying eigenstates and favor (suppress) resonant excitation transport to the sink. While for small electric fields (compared with the average level spacing) one can achieve an enhancement of the quantum coherent transport by choosing an optimal field strength, large electric fields induce the localization of all eigenstates that strongly suppresses any transport; efficient transport can thus be restored only on increasing dephasing (incoherent transport). Simultaneously, a sufficiently strong electric field increases the gap between the largest decay width and the average decay width of all other (subradiant) states induced by the opening to the continuum of states \cite{Chavez2019}. However, in contrast to the superradiance from the coupling to the sink (i.e. ST in the opening) \cite{Kropf2019}, this does not lead to fast quantum coherent transport as we will demonstrate below.

We provide analytical formulae in the limits of low and large electric field and/or dephasing. For large dephasing, transport is diffusive-like. For large electric field, using Leegwater theory \cite{Leegwater1996}, we have found that the  average transfer time is proportional to the square of the potential gradient. Thus, in this region  transport efficiency decreases with  an increasing electric field. In the end we provide a mean to describe the charge current using average transfer time and compute the conductance of the system.

Overall, we show that while a moderate intrinsic electric field can indeed help transport even in the quantum coherent regime, if it is too strong, transport can only occur through classical diffusion which is a much slower and inefficient process.


\section{The Model and the relevant observables}

The starting point of our approach, already introduced in Ref.~\cite{Kropf2019}, is to model the interlayer transport by the single-excitation manifold of a one dimensional N-site chain of two-level systems, which represent the photo-excitation across the gap of the TMO. The many-body interactions, which lead to the effective broadening of the electronic levels, are accounted for by a dephasing term that tends to suppress quantum transport. In order to obtain a photo-current through the multi-layered TMO along the perpendicular axis which we model as a one-dimensional chain, we need to break the symmetry so that the photo-generated charges migrate to one end of the chain. We achieve this by attaching a sink made of a metallic material with a lower Fermi-level to one end of the chain, and assuming hard boundary on the other end from a material with a larger Fermi-level. The addition of a constant electric field potential is in itself not symmetry-breaking in the quantum regime, and thus cannot be the sole drive of the current. However, as we shall see below, a ladder pointing towards the sink can help transport by giving a finite momentum to an initial Gaussian state which then experiences less interference. 

More precisely, we model the transport of an initial photo-excitation $\ket{\psi_0}$ in the single-excitation manifold by a one-dimensional chain of $N$ sites $\ket{j}$ with nearest-neighbour-coupling $\Omega$ attached to a sink $\ket{N}$ (tight-binding model) as illustrated in Fig.~\ref{fig:Fig1}. In addition, we consider a constant (potential) energy shift $E_0 = V / N$ induced by an external or internal constant electric potential $V$ over the chain length $N$ that for simplicity we define always positive $E_0>0$ (negative potentials are denoted by $-E_0$). The Hamiltonian of the system reads
\begin{align}
	\hat H = E_0 \sum_{j=1}^N j \ketbra{j}{j} - \Omega\sum_{j=1}^{N-1} \left(\ketbra{j}{j+1} + \ketbra{j+1}{j} \right) .	
\end{align}
In all the manuscript we will use $\Omega$ as the reference energy scale. The coupling to the sink is described within the Lindblad master equation formalism \cite{Gurvitz1996} and is described by the rate $ \gamma_{out}$ and the corresponding Lindblad operator, 
\begin{align}
	  \hat L_{out} = \ketbra{0}{N}.
\end{align}
The decay rate to the sink is characterized by the tunneling rate $\Omega_{N}$ from site $N$ to the sink and by the density of states $\sigma$ in the sink. In the wide-band limit, the density of states is constant and $\gamma_{out} \sim \Omega_{N}^2 \sigma$ \cite{Gurvitz1996,Giusteri2015}. 

Any additional decoherence from the coupling to other environmental degrees of freedom such as phonons or other excitations is effectively modelled\cite{Haken1973,Breuer2002} by dephasing operators $\hat L_j^\phi = \ketbra{j}{j},\,j=1,\ldots,N$ with a constant, homogeneous rate $\gamma_\phi$. The time-evolution is obtained by solving the corresponding Lindblad master equation 
\begin{align} 
	\frac{d}{dt}\hat\rho =& -\frac{i}{\hbar} \left[\hat H,\hat\rho\right] + \gamma_\phi\sum_{j=1}^N  {\hat L_j^\phi} \hat\rho {\hat L_j^\phi}^\dagger -\frac{1}{2}\left\{ {\hat L_j^\phi}^\dagger  {\hat L_j^\phi}, \hat\rho\right\} \nonumber \\
	& +\gamma_{out} \left(\hat L_{out} \hat\rho \hat L_{out} -\frac{1}{2}\left\{\hat L_{out}^\dagger \hat L_{out}, \hat\rho\right\}\right).\label{eq:Lindblad}
\end{align}
For a finite-size chain, any initial excitation asymptotically reaches the sink. Hence, as a figure of merit for the transport efficiency we consider the  average transfer time to the sink \cite{Kropf2019} defined as
\begin{align}\label{eq:taudef}
	\tau := \frac{\gamma_{out}}{\hbar \eta}\int_0^\infty t \matelem{N}{\hat \rho (t)}{N} dt,
\end{align}
where $\eta =1$ is the asymptotic probability to be on the sink. Our goal is to find parameters to optimize the transport to the sink, which means to minimize the average transfer time $\tau$. 

In the absence of dephasing ($\gamma_\phi = 0$) we can also use the non-Hermitian Hamiltonian  formalism \cite{Celardo2009} which will be useful for later analysis. In this case, one  considers only the Hilbert space of the sites inside of the chain. The decay to the sink is described by an imaginary term $W = -\frac{i\gamma_{out}}{2}\ketbra{N}{N}$. The total (non-Hermitian) Hamiltonian then reads
\begin{align}\label{eq:HnH}
	\hat {\cal H}  = \hat H -\frac{i\gamma_{out}}{2}\ketbra{N}{N}.
\end{align}
Then, denoting as $\ket{E_\alpha}$ and $\bbra{E_\alpha}$ the right and left eigenvectors of $\hat {\cal H} = \sum_{\alpha} E_\alpha \ketbbra{E_\alpha}{E_\alpha}$ , we can compute the time-evolution of an initial state $\ket{\psi_0}$ as
\begin{align*}
	\ket{\psi(t)}  = \sum_{\alpha= 1}^N e^{-\frac{it}{\hbar}E_\alpha} \ket{E_\alpha}\bbra{E_\alpha}\psi_0\rangle.
\end{align*}
and from that the average transfer time to the sink
\begin{align}
	\tau 
	=& \gamma_{out} \int_0^\infty t |\langle N | \psi(t)\rangle |^2  dt \\
 =& \sum_{\alpha,\beta = 1}^N  \langle N\ket{E_\alpha} \left(\langle N\ket{E_\beta}\right)^*
\bbra{E_\alpha}\psi_0\rangle \left( \bbra{E_\beta}\psi_0\rangle\right)^* I_{\alpha,\beta}. \nonumber
\end{align}
where the integral $I_{\alpha,\beta}$ can be solved by integration by part and yields
\begin{align}
	I_{\alpha,\beta} =  \int_0^\infty t e^{-\frac{it}{\hbar}(E_\alpha-E_\beta^*)} dt = \frac{-\hbar}{(E_\alpha-E_\beta^*)^2}.
\label{eq:tauNH}
\end{align}
Thus, as expected, the average transfer time $\tau$ is fully characterized by the eigenvalues and eigenvectors of $\hat{\cal H}$.

\section{The Non-hermitian approach : widths, eigenstates and the Superradiant Transition}

Before investigating the dynamics, let us analyze the behavior of eigenvalues and eigenfunction of the  non-Hermitian Hamiltonian Eq.~(\ref{eq:HnH}). Exact diagonalization produces complex eigenvalues $E_\alpha = {\cal E}_\alpha -i\Gamma_\alpha/2$ where $\Gamma_\alpha$ are the decay widths, responsible for the probability decay to the sink. 

\begin{figure}
\centering
	\includegraphics[width =0.5\textwidth]{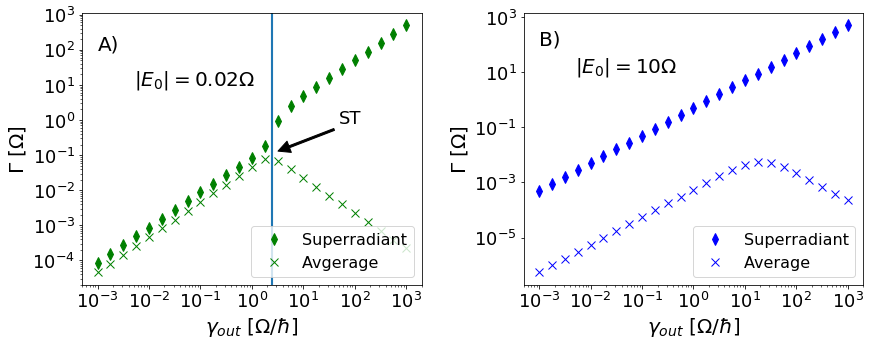}
		\caption{[Color online] Decay width ($\Gamma_\alpha$ of $\hat{\cal H}$, Eq.~\eqref{eq:HnH})
for the Superradiant state (full symbols) and for the average of all other subradiant states (crosses)
 as a  function of $\gamma_{out}$ for a chain of $N=10$ sites. A) Below the critical field $|E_0| = 0.2\Omega < \tilde E_0 \approx 0.56\Omega$, Eq.~\eqref{eq:CriticalE0},   B) Above the critical value, $|E_0| = 10\Omega>\tilde E_0\approx 0.56 \Omega $.
}
		\label{fig:Fig22}
\end{figure}

In Fig.~\ref{fig:Fig22} the largest decay width and the average decay width of all other states are shown as a function of the coupling to the sink $\gamma_{out}$ for two different values of the electric field (left and right panels). In both cases the picture is very close to that considered in \cite{Celardo2009,Celardo2012} : at some critical value of the opening  strength $\gamma_{out} \simeq  \gamma^{ST}$, one width increases with $\gamma_{out}$ while the average value of the others decreases. This has been called Superradiant Transition (ST), meaning that the state with the largest width (i.e. fastest probability to decay) is "Superradiant", while the others, whose average width decreases for large opening strength become subradiant \cite{Chavez2019}. As one can see from a comparison between the two panels, an increase of almost three orders of magnitude in the electric field increases the width separation between the superradiant and the average for small values of $\gamma_{out}$, but does not change qualitatively the overall picture. From this point of view, we may say that the ST is robust to the application of an electric field as far as the sharp opening of a gap between the largest and the average width as a function of the coupling to the sink $\gamma_{out}$ is concerned.

\begin{figure}
\centering
	\includegraphics[width =0.5\textwidth]{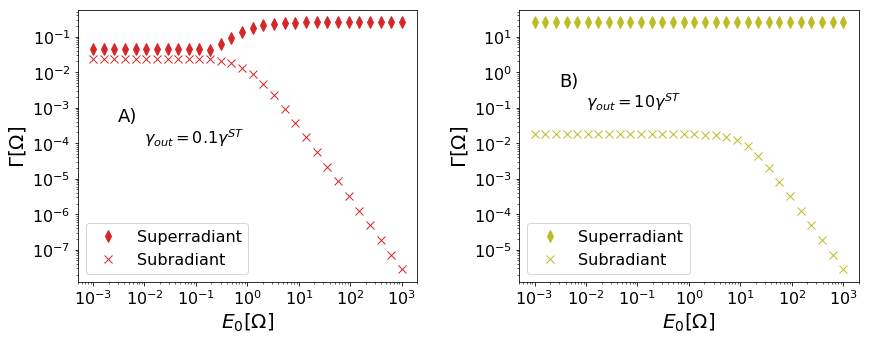}
		\caption{[Color online] Decay width ($\Gamma_\alpha$ of $\hat{\cal H}$, Eq.~\eqref{eq:HnH}) 
for the Superradiant state (full symbols) and for the average of all other subradiant states (crosses)
 as a  function of $E_0$ for a chain of $N=10$ sites.
 A) The opening $\gamma_{out}=0.1 \gamma^{ST}$ is below the ST,   B) The opening $\gamma_{out}=10 \gamma^{ST}$ is above the ST.
}
	\label{fig:Fig222}
\end{figure}

One may wonder if an analogous transition occurs at fixed opening (below and above the ST) on changing the electric field. The answer to this question is in Fig.~\ref{fig:Fig222}. As one can see both below and above the ST, on increasing the value of the electric field $E_0$ a further width separation occurs with the sharp increase of the gap between the largest width $\Gamma_{max}$ (superradiant) and the average width over all others states $\langle\Gamma\rangle$ (subradiant). Quite naturally, since $\gamma_{out}$ is now fixed, at variance with the previous case in which we fixed the value of $E_0$, the total probability to decay (sum over all widths) is now a constant. 

\begin{figure}
\centering
	\includegraphics[width =0.5\textwidth]{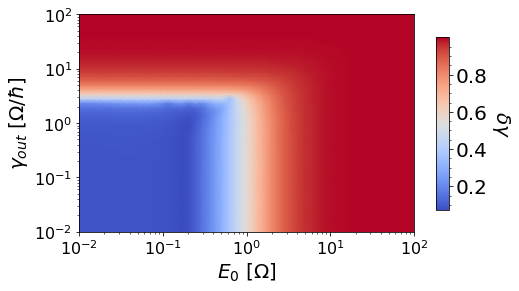}
		\caption{[Color online] Normalized gap $\delta\gamma$, see Eq.~(\ref{eq:dg}) as a function of both 
$\gamma_{out}$ and $E_0$ for a chain of $N=10$ sites which shows the Superradiant border (white area).
}
	\label{fig:Fig22a}
\end{figure}

In order to make the two cases more "symmetric", let us consider the normalized gap, defined as 
\begin{align}\label{eq:dg}
	\delta \gamma = \frac{\Gamma_{max}-\langle\Gamma\rangle}{\gamma_{out}}
\end{align}
that we plot in Fig.~\ref{fig:Fig22a} as a function of both $\gamma_{out}$ and $E_0$.

As one can see, the opening of a gap appears now to be symmetric in both variables $\gamma_{out}$ and $E_0$ and we may speak about a Superradiant border (in the figure, roughly speaking the separation between the blue rectangle and the red border). Even if, from the point of view of the gap opening in the widths, $\gamma_{out}$ and $E_0$ seem to behave in the same they will affect transport properties in a very different way.

\begin{figure}[t]
\centering
	\includegraphics[width =0.5\textwidth]{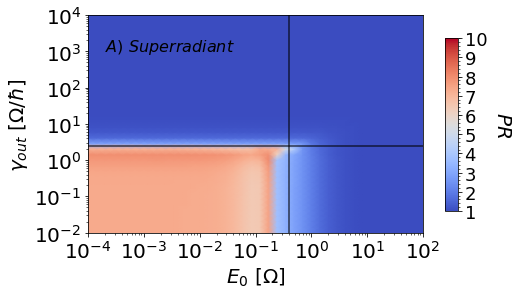}
       \includegraphics[width =0.5\textwidth]{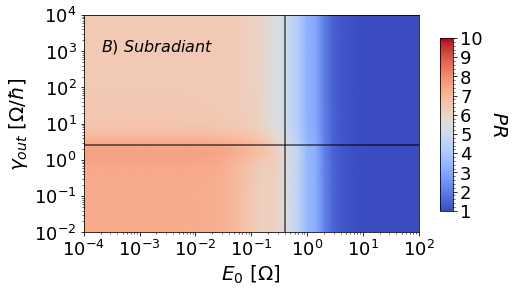}
		\caption{[Color online] Participation ratio for the Superradiant state (panel A) and average participation ratio of all subradiant states (panel B) , as a function of both $E_0$ and $\gamma_{out}$ for a chain of $N=10$ sites. Vertical  lines represent the average level spacing $4\Omega/N$, horizontal ones the ST transition. 
}
	\label{fig:Fig22b}
\end{figure}

In order to better address this point, let us consider another important quantity that affects transport properties : the degree of localization of the eigenstates. One standard measure of their localization length is the so-called participation ratio $$PR_\alpha = 1/\sum_k |\langle k | E_\alpha \rangle|^4,$$ where $\ket{k}$ is the site basis and $1 \leq PR_\alpha \leq N$ measures approximatively the "number of significantly occupied sites" by the eigenstate $\ket{E_\alpha}$ in a chain of length $N$. In similar way to what we have done for the widths, let us consider the participation ratio of the Superradiant state and the  average participation ratio of all other subradiant states as a function of both $E_0$ and $\gamma_{out}$. Results are shown in Fig.~\ref{fig:Fig22b}. As one can see the difference between the transition in $E_0$ and $\gamma_{out}$ is very different -- while the former induces a localization of all eigenstates, the latter produces the localization of only the superradiant eigenstate (at the sink $N$).

This of course will have dramatic consequences on the transport properties, specially regarding the dependence on the initial state as will be studied in the next section.

\section{Results on transport : the average transfer time}

We start with a numerical study of the average transfer time followed by analytical calculations for different limiting behaviour. The average transfer time is computed by explicit integration of $\tau$ and numerical diagonalization of the Liouvillian associated with the Lindblad equation \eqref{eq:Lindblad} (c.f. Appendix \ref{app:Liouvillian}). In order to compute $\tau$, we use the Python packages \textit{Qutip} \cite{Johansson2013} and \textit{numpy} for the numerical diagonalization of the Liouvillian.

We consider a chain of $N=10$ sites, which was shown to be short enough to expect quantum coherent transport in TMOs heterostructures~\cite{Kropf2019}.  The initial photo-excitation is modelled by a generic Gaussian initial state that we choose to be localized on site $n_0=3$, which is far enough from the sink to obtain meaningful conclusion about transport, to have width $\Delta_0 = 1$ (in lattice size units) and no initial momentum $k_0 =0$,
\begin{align}\label{eq:gauss3}
	\ket{\psi_0} = \sum_{n=1}^N e^{-ik_0 n}e^{-\frac{(n-n_0)^2}{4 \Delta_0^2}}\ket{n}
\end{align}
After the general discussion we shall comment on the impact on the average transfer time of longer chains and different initial states.

\subsection{Negligible dephasing $\gamma_\phi \approx 0$.}
\begin{figure}
\centering
	\includegraphics[width =0.5\textwidth]{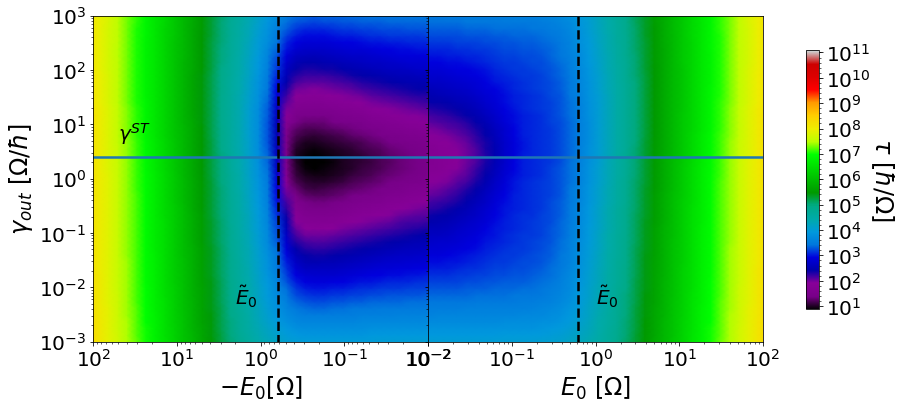}
		\caption{[Color online] Average transfer time in function of the electric field gradient $E_0$ and the coupling rate $\gamma_{out}$ to the sink at negligible dephasing $\gamma_{\phi} = 0.001$. Optimal transport occurs in the black region. This region is characterized by an optimal coupling to the sink $\gamma_{out} = gamma^{ST}$ (horizontal line), and by an optimal electric field. The regime where quantum coherent transport occurs is delimited by the two dashed vertical lines (see Eq.~\eqref{eq:CriticalE0} and discussion below). Please note the axes jump at $|E_0|=0.01$. The initial state is a Gaussian state Eq.~\ref{eq:gauss3} with $\Delta_0=0, n_0 = 3, k_0 = 0$. 
}
		\label{fig:Fig2}
\end{figure}
The average transfer time as a  function of the electric field and the coupling to the sink at vanishing dephasing $\gamma_\phi \approx 0$ is shown in Fig.~\ref{fig:Fig2} for a Gaussian initial state (c.f. Eq.~\eqref{eq:gauss3}). As one can see, there is both an optimal value of the opening $\gamma_{out}$ and of the electric field $E_0$ at which the average transfer time is minimized. As we will show later, the optimum in $E_0$ is strongly dependent on the initial state. For instance, for any localized initial state the optimal field is $E_0= 0$, see Section \ref{sec:initialstate}. 

For an electric field gradient smaller than  some critical value $\tilde{E}_0$  (indicated by vertical dashed lines)  the average transfer time is always minimized at the ST\cite{Celardo2012,Kropf2019}, $\gamma_{out} = \gamma^\textrm{ST} \approx 2\Omega$ . In this figure, as vertical dashed line, we indicate  the ``critical'' electric potential $\tilde E_0 = 4\sqrt{2}\Omega/N$ derived below, see Eq.~\eqref{eq:CriticalE0},  which turns out to be on the same order of the average level spacing.
With ``critical'' we mean that  any electric field exceeding such value is detrimental for transport. 

Let us interpret physically this result. For small electric fields, $|E_0| < \tilde E_0$,  the transport is coherent, and is optimized at the ST and  and at a particular value of the electric field. For a generic initial state which is a uniform superposition over all eigenstates the probability to reach the sink in a shorter time increases on increasing the coupling to the sink. Nevertheless, on increasing $\gamma_{out}$ above the ST  one state, the superradiant one, decays faster than any other and, at the same time starts to be localized at the sink. This means that all other subradiant states will have a negligible overlap with the sink (due to the eigenfunction normalization) thus producing an effective barrier to transport. In other words the transport will be optimized just in the middle of these two situations (a large enough shared width and a low enough localization to the sink) which is exactly the ST.  On the other hand, a further increase of the electric field $E_0$, larger than the average level spacing will produce a localization of all eigenstates thus producing a complete blocking of transport. In this regime, changing the value of $\gamma_{out}$ has practically no effect on the average transfer time for initial states that do not strongly overlap with the superradiant state as can be seen in Fig.~\ref{fig:Fig2}. 

Note also that the average transfer time at a given $|E_0|<\tilde E_0$ is minimized at the optimal coupling to the sink $\gamma^{ST}$, but the average transport time $\tau$ is lower for $E_0 <0$ than for $E_0>0$. This is because a potential gradient towards the sink favors a shorter transfer time as will be discussed in details below. 

This is quite important since it shows that it makes no sense to look for an optimal electric field considering only the widths, but not the initial conditions. Indeed from Fig.~\ref{fig:Fig2} it is clear that an "optimal" electric field also exists but this cannot be related to the ST transition in the widths, which occurs also for negative $E_0$ (since the widths do not depend on the sign of $E_0$) while for our initial Gaussian state transport is optimized only for negative $E_0$ and not for positive. This suggests that the choice of the initial state is crucial for understanding the optimal transport conditions.

\subsection{Adding dephasing : optimal electric field gradient}
\begin{figure}[t]
\centering
	\includegraphics[width =0.5\textwidth]{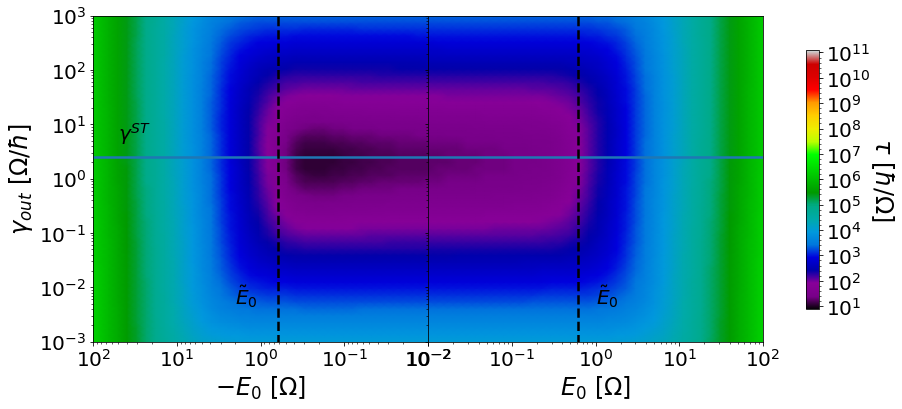}
        \includegraphics[width =0.5\textwidth]{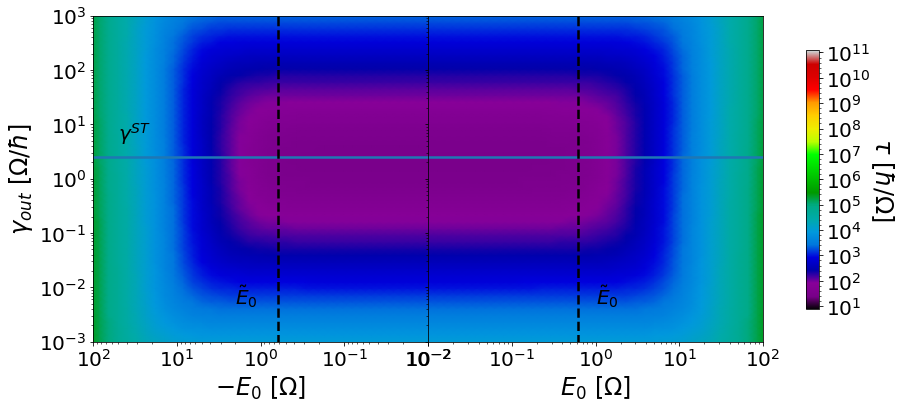}
        \includegraphics[width =0.5\textwidth]{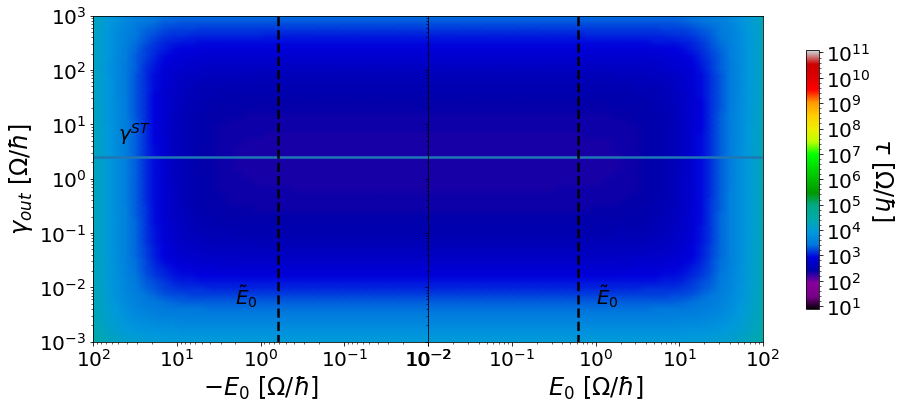}
                 \caption{[Color online] Average transfer time as a  function of the electric field gradient $E_0$ and the coupling rate $\gamma_{out}$ to the sink for small dephasing $\gamma_{\phi} = 0.1$ (top panel), 
moderate dephasing $\gamma_\phi=1$ (central panel) and large dephasing $\gamma_\phi = 10$ (bottom panel). Vertical and horizontal lines and the initial state are the same as in Fig.~\ref{fig:Fig2}.
}
		\label{fig:deph}
\end{figure}
\begin{figure}
\centering
	\includegraphics[width =0.5\textwidth]{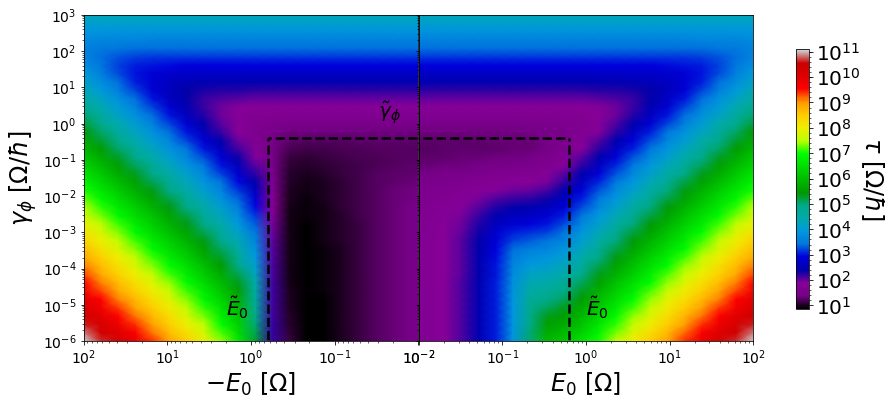}
		\caption{[Color online] Average transfer time in function of the electric potential $E_0$ and the dephasing rate $\gamma_\phi$ at the superradiant transition $\gamma_{out} = \gamma^{ST}$ for a chain of $N=10$ sites. At large electric fields, (i.e., $>\tilde E_0$) dephasing is required for transport and is slow. For low electric fields (i.e., $<\tilde E_0$), just below the critical dephasing $\tilde \gamma_\phi$ signalling the transition to the diffusive-like transport, there is an optimal dephasing value. At low dephasing ($\gamma_\phi < \tilde \gamma_\phi$), transport is optimal at a finite value of the electric field gradient smaller than the critical value $\tilde E_0$, Eq.~\eqref{eq:CriticalE0}. Note the axes jump at $|E_0|=0.01$. The initial state is a Gaussian state Eq.~\ref{eq:gauss3} with $\Delta_0=0, n_0 = 3, k_0 = 0$. 
}
		\label{fig:Fig3}
\end{figure}
Similar plots, as that presented in Fig.~\ref{fig:Fig2}  but for increasing   values of the dephasing $\gamma_\phi$
are shown in Fig.~\ref{fig:deph}. While the top panel indicates that the picture for zero dephasing persists even for larger dephasing strength $\gamma_\phi = 0.1$, the other panels show that a further increase of the  dephasing blurs everything. In a sense transport is again "slightly optimized" but this happens for a wide region of the parameters $E_0$ and $\gamma_{out}$. We can see that transport is, in any case,  always optimized when the opening strength is set  at the critical value $\gamma^{ST}$.

In order to  estimate more precisely the effect of the dephasing, let us thus consider the transport exactly at the ST (i.e. $\gamma_{out} = \gamma^{ST}$)  and study the transport time as a function of both $E_0$ and $\gamma_\phi$. Results are  shown in Fig.~\ref{fig:Fig3}. The black region in this figure shows that transport is optimized for suitable values of the dephasing and electric field strengths. On the other hand, for large electric fields $|E_0|>\tilde E_0$ transport is slow and increasing $E_0$ is always detrimental to transport. This is due to the localization of the eigenstates and so, the diffusive-like transport can only happen in the presence of dephasing. 

For large dephasing, the situation is analogous and transport is diffusive-like (at least in an infinite chain). The combination of both large dephasing and large electric field gives rise to dephasing-assisted transport~\cite{Plenio2008}, i.e., when dephasing is of the same order of the energy separation between levels, energy fluctuations can support transport by inducing resonances between levels. As shown in Fig.~\ref{fig:Fig3}, for any fixed value of the electric field gradient
$|E_0| > \tilde E_0$ (indicated by vertical dashed lines) an optimal value of $\gamma_\phi$ exists minimizing the transport time by the dephasing-assisted transport mechanism. 

Let us now turn our attention to the most interesting regime in which the electric field gradient is below the critical value $|E_0|\lesssim\tilde E_0$ and the dephasing is small enough as compared to the average level spacing to allow for quantum coherent transport. Then the most important parameter to optimize transport is to tune the coupling to the sink at the ST  as shown for $\gamma_\phi = 0, E_0\lesssim \tilde E_0$ in Fig.~\ref{fig:Fig2}, and as was established in \cite{Kropf2019} for $E_0 =0, \gamma_\phi \lesssim \tilde \gamma_\phi \approx 4\Omega/N$. In addition we find two additional optimal parameters as shown in Fig.~\ref{fig:Fig3}~: a wide interval of optimal dephasing rates which is independent on the electric field values, i.e., it is not related to dephasing-assisted transport, and an optimal electric field. While the former has a rather small impact on the average transfer time and is essentially due to a small broadening of the energy levels, the latter can shorten the average transfer time by up to an order of magnitude. Since we are in the coherent transport regime, transport is optimized when the overlap between the initial state and the radiant states is maximized. We recall that at the ST the average decay of every eigenstates is maximized. This guarantees an overall efficient transport for any initial state, even when not localized on the superradiant state. In addition, the electric field can fine-tune the best overlap. The main effect of a non-vanishing $E_0$ is to modify the eigenstates, while conserving the overall structure of the eigenvalues. While the eigenvalues are insensitive to the direction of the electric gradient  this is not the case for the eigenstates and explains the asymmetry in the average transfer time shown in Fig.~\ref{fig:Fig3}. Indeed, $\tau$ is smaller for an electric gradient towards the sink, i.e., $E_0 < 0$.

\begin{figure}
\centering
	\includegraphics[width =0.5\textwidth]{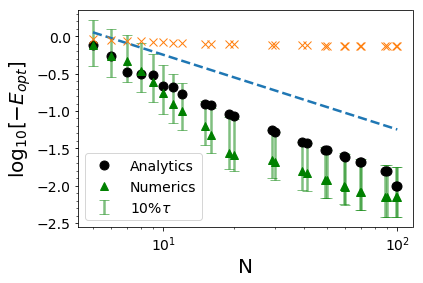}
		\caption{Electric field gradient at which the average transfer time is minimized as a function of the chains length, numerical grid search (triangles) and analytic estimate Eq.~\eqref{eq:Eopt} (dots), at the superradiant transition $\gamma_{out} = \gamma^{ST}$ and small dephasing $\gamma_\phi = 10^{-6}$. The errorbar indicates the electric field gradient with $10\%$ variations in $\tau$. The initial state is a Gaussian state Eq.~\eqref{eq:gauss3} with $\Delta_0 = 1$, $n_0 =3$ and $k_0=0$ for all $N$. As a visual guide the critical field $\tilde E_0$ (blue dashed) above which localized eigenstates suppress coherent transport, and the field (orange crosses) at which the normalized gap between the super- and sub-radiant states ($\delta \gamma =0.5$, see Eq.~\eqref{eq:dg}) opens are shown.
}
		\label{fig:Eopt}
\end{figure}
Having in mind the discussion above, we propose here below a phenomenological expression able to capture the electric field at which transport time is minimized (given that $\gamma_{out} = \gamma^{ST}$).
The idea is that transport will be optimal when the initial state is overlapped with many eigenstates whose widths are significantly different from zero. Therefore we will expect an optimal electric field gradient by minimizing (and not maximizing since $\textrm{Im}[E_k]<0$ $\forall k$)

\begin{align}\label{eq:Eopt}
	 E_0^{opt} = \underset{E_0}{\min}\sum_{k=1}^N \textrm{Im}[E_k] |\braket{\psi_0}{E_k}|,
\end{align} 

Since from Figs.\ref{fig:Fig2},\ref{fig:Fig3} the optimal transport time occuring at small dephasing and at the ST is weakly dependent on both values $\gamma_\phi$ and $\gamma_{out}$, we check the validity
of Eq.~\ref{eq:Eopt}, by changing the length of the chain $N$ keeping the initial state Eq.~\eqref{eq:gauss3}. Results are shown in Fig.~\ref{fig:Eopt}. We plot the results from a numerical grid search for the $E_0$ that minimizes $\tau$ with an error bar given by the $\pm E_0$ values corresponding to 10\% variation in $\tau$ around the optimal value. This we do since the minimal times are usually characterized by large plateaus that can mask artificially the predictions, see for instance Fig.~\ref{fig:deph}. As one can see, the analytical results fit well the numerical ones for small chain lengths, while some deviations occur at larger $N$ values, but fit within the $10\%$ plateau. The case of long chain actually is quite off from our main considerations concerning quantum coherent transport, being even in the presence of small dephasing typically characterized by ``classical'' diffusion.

\begin{figure}
\centering
	\includegraphics[width =0.5\textwidth]{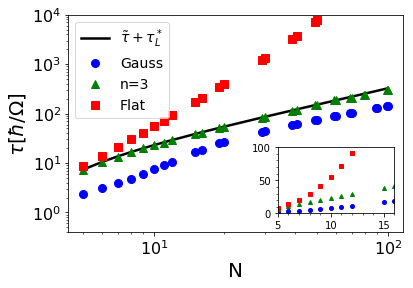}
		\caption{Minimal $\tau$ as a function of the number of sites $N$ for an initially localized state  
 $\ket{3}$ (green triangles), the Gaussian state Eq.~\eqref{eq:gauss3} (blue circles) and a flat state $1/(N-1)\sum_{j=1}^{N-1} \ket{j}$ (red squares) at the optimal electric field for the given initial state and chain length ($E_{opt}$ from Fig.~\ref{fig:Eopt} (green) and $E_0 = 0$ (red, blue)). Dephasing is negligeable $\gamma_\phi = 10^{-6}$ and the opening is set at the ST : $\gamma_{out}=\gamma^{ST}$. The fastest transport is achieved for the Gaussian wavepacket at the electric field shown in Fig.~\ref{fig:Eopt}. The heuristic equation Eq.~\eqref{eq:heuristicfinal} reproduces the numerics for the localized state. Inset : the same in normal scale on both axis and for small lengths.
}
		\label{fig:Eopt2}
\end{figure}

\begin{figure*}
	\includegraphics[width = \textwidth]{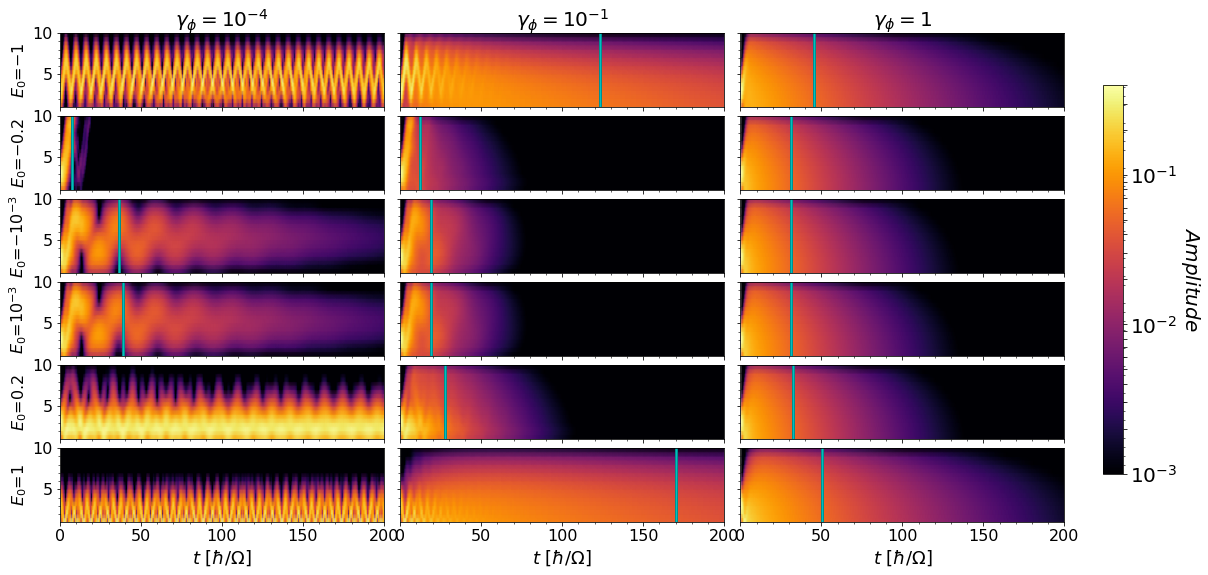}
	\caption{Time-evolution of the probabilities $|\langle j | \psi(t) \rangle |^2 $ for different values of the electric field gradients (from top to bottom $E_0 = [-1, -0.2, -0.001, 0.001,0.2,1]$) and of the dephasing (From left to right $\gamma_\phi= [0.0001, 0.1, 1]$) for a chain of $N=10$ sites. Vertical cyan line indicates the average transfer time $\tau$. At the optimal value of the electric field (i.e, $E_0 \approx -0.2$, second line from the top), self-interference is  minimized. Moderate dephasing (central column) also helps transport by destroying self-interferences without making the transport diffusive-like. The initial state is a Gaussian state Eq.~\ref{eq:gauss3} with $\Delta_0=0, n_0 = 3, k_0 = 0$. 
}
\label{fig:Fig7}
\end{figure*}

\subsection{Changing the initial state}\label{sec:initialstate}
Under optimal conditions, the Gaussian state is transferred to the sink faster than a completely localized ($\ket{\psi_0} = \ket{3}$) or delocalized state ($\ket{\psi_0} = 1/\sqrt{N-1}\sum_{j=1}^{N-1} \ket{j}$), as shown in Fig.~\ref{fig:Eopt2}. It is also important to observe that for an initial state  localized on a single site  a non-vanishing electric field is always detrimental, see for details \ref{app:Localized}. Indeed, in this case   one cannot avoid self-interference, and thus the electric field cannot enhance transport. Localized states are completely delocalized in momentum space and thus always have a component which will be driven away from the sink. However, we remark that $\tau$ can be efficiently minimized by the introduction of dephasing. It was shown that this is related to the diffusion of the state from low momentum component to larger momentum components\cite{Li2015} (also termed "momentum rejuvenation").

\subsection{A dynamical point of view}

Another point-of-view on the optimization of the average transfer time by non-vanishing electric field and dephasing in the coherent transport regime is obtained in the temporal regime. In Fig.~\ref{fig:Fig7} we show the time-evolution of the site populations  for different values of $\gamma_\phi$ and $E_0$ at the superradiant transition. In the absence of dephasing and electric field, the initial Gaussian wave-packet will first spread ballistically. Then, since the wavepacket starts close to the edge of the chain which is described by a hard wall, part of the wavepacket will be reflected and will begin to interfere with itself. These interferences then prevent a smooth movement of the particle towards the sink and leads to a large transport time. Introducing a negative electric field then allows to give an initial momentum away from the wall, thus minimizing the self-interference and optimizing the transport. Consequently, positive electric field never lower $\tau$ (c.f. Fig.~\ref{fig:Fig3}), and particles with initial momentum towards the sink do not benefit as much from the tuning of $E_0$. On the other hand, introducing dephasing does not provide a directionality, but suppresses any coherence, and thus also unwanted interferences. In both cases there is an optimal value favoring transport. Introducing a too large dephasing leads to slow diffusive-like transport, and a too large electric field leads to a localization of all eigenstates, both of which eventually kill the quantum coherent transport.

\subsection{Comparison with disorder}

\begin{figure}
\centering
	\includegraphics[width =0.5\textwidth]{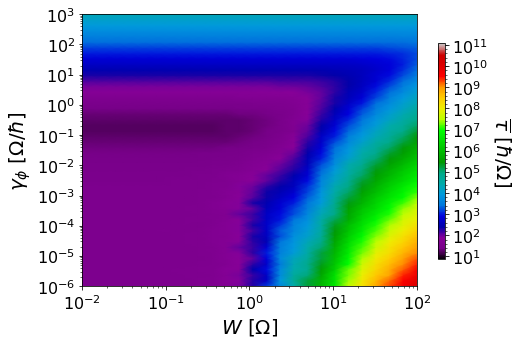}
		\caption{Average transfer time in function of the dephasing $\gamma_\phi$ and the disorder strength $W$ at the ST $\gamma_{out} = \gamma^{ST}$ for a chain of $N=10$ sites with on-site Anderson-type disorder of width $W$ averaged over $1000$ realizations of the disorder. Contrarily to the case of the electric field, there is no optimal disorder value. The optimal dephasing value at low disorder is however still present. The initial state is a Gaussian state Eq.~\ref{eq:gauss3} with $\Delta_0=0, n_0 = 3, k_0 = 0$.}
		\label{fig:Disorder}
\end{figure}

It is clear from both the spectral as well as the time analysis that the electric field does not act as a static (diagonal) disorder in the quantum coherent regime. On the other hand, once $E_0$ is large enough to localize the eigenstates, it is similar to Anderson-type disorder as shown in Fig.~\ref{fig:Disorder} (see also the r.h.s. of Fig.~\ref{fig:Fig3}). Indeed, if we set $E_0 = 0$ and introduce independent uniform disorder $\epsilon_j$ sampled from $[-W/2,W/2]$ on each site $j$, the ensemble-averaged  $\overline \tau$ never gets smaller with increasing disorder $W$. In other words, there is no optimal value of disorder, as opposed to the existence of an optimal value of the electric field. For a disorder $W \gtrsim 10\Omega/\sqrt{N}$ such that the localization length is expected to be shorter than the chain's length \cite{Izrailev1998}, transport is strongly suppressed and requires non-vanishing dephasing. Therefore, in this case, the disorder has a similar impact on the average transfer time as electric field gradients $|E_0| \gtrsim \tilde E_0$, as can be seen by comparing Figs.~\ref{fig:Disorder} and \ref{fig:Fig3}. For values $W \lesssim 10\Omega/\sqrt{N}$, the effect of the disorder appears similar to $0<E_0<\tilde E_0$. 

\section{Analytical results}

\subsection{Perturbation theory}

We can obtain a good analytical description of the average transfer time in the limits of small coupling to the sink $\gamma_{out}$ using perturbation theory, and for large dephasing or/and electric field using a heuristic approach. We start by a non-degenerate perturbation expansion in $\gamma_{out}$ of the eigenvalues and eigenstates of the non-Hermitian Hamiltonians, Eq.~\eqref{eq:HnH}, for $E_0=0$, $\gamma_\phi =0$. To obtain the lowest order of $\tau$, \eqref{eq:tauNH}, in $\gamma_{out}$, we need the second-order expansion of the eigenvalues 
\begin{align}
	E_\alpha =& 2t \co{\piN{\alpha}} -\frac{\gamma_{out}}{(N+1)^2}\sum_{\xi \neq \alpha} \frac{S_{\alpha\xi}^2(N,N)}{4tQ_{\alpha\xi}} \nonumber\\
	&-\frac{\gamma_{out}^2}{(N+1)^2}\sum_{\xi \neq \alpha} \frac{S_{\alpha\xi}^2(N,N)}{4tQ_{\alpha\xi}} + O(\gamma_{out}^3)
\end{align}
due to the $1/(E_\alpha-E_\beta^*)^2$ term in $\tau$. Here we defined $Q_{\alpha\xi} := \frac{1}{2}\left[\co{\piN{\alpha}}-\co{\piN{\xi}}\right]$ and $S_{\alpha\xi}(j,k) := \si{\piN{\alpha j}}\si{\piN{\xi k}}$ to lighten the notation. We also need the unperturbed part of the eigenvectors
\begin{align}
	\ket{E_\alpha} = \sqrt{\frac{2}{N+1}}\sum_{k=1}^N\si{\piN{\alpha k}} \ket{k} + O(\gamma_{out}).
\end{align}
Then, we obtain the lowest-order in $\gamma_{out}$ contribution to the average transfer time from Eq.~\eqref{eq:tauNH}
\begin{align}\label{eq:tau0}
	\tilde\tau = \frac{\hbar}{\gamma_{out}}\sum_{j,k=1}^N\rho_{jk}(0) \left(\sum_{\alpha = 1}^N \frac{\si{\piN{\alpha j}}\si{\piN{\alpha k}}}{\si{\piN{\alpha N}}^2}\right).
\end{align}
For an initial state $\psi_0 = \ket{n}$ localized on site $n$, this simplifies to 
\begin{align}\label{eq:tauperturbn}
	\tilde\tau = \hbar\frac{n(N-n+1)}{\gamma_{out}},
\end{align}
which for $n=1$ is consistent with the formula described in \cite{Kropf2019, Zhang2017a} and perfectly agrees with the numerical simulations.

\subsection{Beyond perturbation theory : a phenomenological approach}

We have found that for $E_0 = 0$, $\gamma_\phi = 0$ and $\ket{\psi_0} = \ket{n}$ the average transfer time is given by
\begin{align}\label{eq:tau1}
	\tau = \hbar \frac{n(N-n+1)}{\gamma_{out}} +\hbar^2\gamma_{out} \frac{n(N-n)}{4\Omega^2},
\end{align}
which is consistent with the findings from \cite{Kropf2019,Zhang2017a} for $n=1$ and converges to the perturbative result, Eq.~\eqref{eq:tauperturbn}, for $\gamma_{out}\to 0$. The average transfer time is thus quite counter-intuitively, minimized for $n=1$, i.e. starting on the site which is furthest from the sink (we omitted the case of an initial state directly attached to the sink). Our interpretation in terms of self-interference applies here: starting at the edge of the chain the excitation can only travel towards the sink and does not bounce on the left wall. At the optimal coupling to the sink (the superradiant transition), $\tau$ is then minimized. While in presence of the electric field it remains in general true that for localized sites transport is optimal starting from the edge of the chain, this is no longer true for non-vanishing dephasing. Then, being closer to the sink becomes advantageous as the transport becomes diffusive-like.

One can use the Leegwaters and Förster theory \cite{Leegwater1996} to derive a heuristic formula for the average transfer time $\tau$. This is valid in the large dephasing limit, in presence of an electric field and for initial states localized on single sites \cite{Zhang2017b}. When starting from the first site, we have 
\begin{align}\label{eq:Leegwaters}
	\tau_L &:= \frac{(N-1)(N-2)}{2\Gamma_F} +\frac{N-1}{\Gamma_L} \\
	\Gamma_F&:= \frac{2\Omega^2\gamma_\phi}{\gamma_\phi^2+E_0^2} \;\;\; ;\;\;\; \Gamma_L := \frac{2\Omega^2(\gamma_\phi+\gamma_{out}/2)}{(\gamma_\phi + \gamma_{out}/2)^2+E_0^2}. \nonumber
\end{align}
The first term in $\tau_L$ describes the diffusion from the first site to the $(N-1)-$th site with the Förster rate \cite{Foerster1948}. The second term is the Leegwater correction to the Förster rate for the last site that is coupled to the single sink. 

When the initial state is localized on a site different from $n=1$, we know that $\tau_L$ must converge to Eq.~\eqref{eq:tau1} for $\gamma_\phi \rightarrow 0$ and $E_0 \rightarrow 0$ (note that the limit should be taken keeping fixed the ratio $E_0 / \gamma_\phi $). Hence, we can correct the second term in Eq.~\eqref{eq:Leegwaters} to $\frac{n(N-n)}{\Gamma_L}$. Furthermore, we should consider that for $n\neq 1$, the diffusion length is reduced to $N-n+1$. Hence, diffusion from site $n$ to site $N-1$ requires $(N-n)(N-n-1)$ steps. We then obtain
\begin{align} \label{eq:leeg}
	\tau_L^* &= \frac{(N-n)(N-n-1)}{2\Gamma_F} +\frac{n(N-n)}{\Gamma_L}\\
	&=\frac{n(N-n)\gamma_{out}}{4\Omega^2} + \frac{(N-n)(N+n-1)}{4\Omega^2}\gamma_\phi \nonumber \\
	& + E_0^2 \frac{N-n}{\Omega^2} \left[\frac{n}{2\gamma_\phi+\gamma_{out}} + \frac{N-n-1}{4\gamma_\phi}\right] \nonumber
\end{align}
Combining the Leegwater rate and the perturbative rate Eq.~\eqref{eq:tau0} we obtain
\begin{align} \label{eq:heuristicfinal}
	\tau = \tilde \tau + \tau_L^*,
\end{align}
which a good estimate for the average time of a state that is initially localized on site $n$ (see Fig.~\ref{fig:Fig7}). This relation can be put in close correspondence with the results found in Ref.~\cite{Kropf2019} where the transport starting from a localized excitation on the first site was studied: even in that case  the strong dephasing solution remains valid for very low dephasing too.. Eq.~\ref{eq:leeg} shows that the average tranport time is quadratic in the electric field gradient in the diffusive regime ($E_0 > \tilde E_0$ and/or $\gamma_\phi > \tilde \gamma_\phi$).

\begin{figure}[]
	\includegraphics[width = 0.5\textwidth]{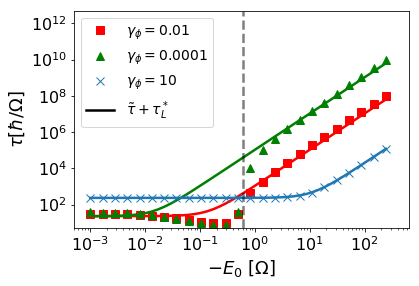}
	\caption{Average transfer time in function of the electric field gradient $E_0$ at the ST    $\gamma_{out} = \gamma^{ST}$ for a chain of $N=10$ sites. Our heuristic model $\tau = \tilde \tau + \tau_L^*$, Eq.~\eqref{eq:heuristicfinal}, (solid lines) describes the average transfer time when self-interference effects inside of the chain are suppressed by large dephasing, $\gamma_\phi > \tilde \gamma_\phi \approx 0.4$, Eq.~\eqref{eq:dephcrit} (blue crosses), or when the eigenstates are localized by the large electric field, i.e., $|E_0|>\tilde E_0 \approx 0.56$, Eq.\eqref{eq:CriticalE0} (indicated with the vertical dashed line). The initial state is a Gaussian state Eq.~\ref{eq:gauss3} with $\Delta_0=0, n_0 = 3, k_0 = 0$. 
}\label{fig:TauEfieldSuper_Leegwaters}
\end{figure}

For an initial Gaussian state centered around site $n$, the heuristic formula Eq.~\eqref{eq:heuristicfinal} works well in the limits of large dephasing and/or large electric field, but fails to capture the quantum transport at the optimal electric field and optimal dephasing as shown in Fig.~\ref{fig:TauEfieldSuper_Leegwaters}. This indicates that the optimization is a genuine quantum effect that cannot be captured by effective rate equations as obtained from Leegwaters and Förster theory.

We can use this to derive the approximate value $\tilde E_0$ of the electric field gradient at which the transition to incoherent transport occurs by the following reasoning. In Fig.~\ref{fig:TauEfieldSuper_Leegwaters}, we compare numerical results with Eq.~\eqref{eq:heuristicfinal}. As one can see the analytics accurately describes the average transfer time for any electric field at large dephasing. On the other hand, at lower dephasing, the formula is not able to describe the data.  We can estimate the value of  dephasing $\tilde\gamma_\phi$ at which this happens directly  from Eq.~\eqref{eq:heuristicfinal}. In order to do that, let us consider the time  $\tau_L^*$ in the absence of electric field and compare the leading order in $N$ contribution of the $\gamma_\phi$ and $\gamma_{out}$ terms. At the ST, $\gamma_{out} \approx 2\Omega$ we find 
\begin{align}\label{eq:dephcrit}
	\tilde \gamma_\phi \approx \frac{4\Omega n}{N+n}
\end{align}
Next, we can find the critical electric field gradient by finding the value of $E_0$ at which the low and large field limit results coincide, i.e., $\tau_L^* = \tilde \tau$, at the ST and at the critical value of the dephasing given in Eq.~(\ref{eq:dephcrit}). We remark that $\tau$ describes the average transfer time when starting from the site $1$ and above the dephasing threshold with high accuracy. When starting from another site, we do not have a close formula for $\tau$  at finite values of the dephasing. Hence, we estimate $\tilde E_0$ for $n=1$ and find
\begin{align}\label{eq:CriticalE0}
	\tilde E_0 \approx 4\frac{\sqrt{2}\Omega}{N}.
\end{align}
which is in good agreement with our  numerical results shown in Figs.~\ref{fig:Fig2},~\ref{fig:Fig3},~\ref{fig:Fig7},~and~\ref{fig:TauEfieldSuper_Leegwaters}. Moreover, it appears intuitive to relate this quantities to the mean average level spacing $\sim 4\Omega/N$. Then it becomes clear that the transition from coherent transport to diffusive-like, dephasing-driven transport occurs roughly when the eigenstates localize under the effect of $E_0$ (c.f. Fig.~\ref{fig:Fig22b}).

\section{Charge current}
For a comparison with TMO materials of technological interest, we derive an appropriate quantity to describe the electric current photo-generated by light excitation across the insulating gap, e.g., the Mott gap in the case of LaVO$_3$.  Since the electric current is defined as the charge per unit time escaping from the collectors, i.e., $\pm e/\tau$, both the electrons and holes injected by photon absorption participate to the total current $I$. In order to obtain a non-vanishing charge current, an asymmetry between the electron and holes transfer time is required. In presence of intrinsic (or extrinsic) electric fields, it is reasonable to assume that the two species will be collected on the two different sides of the chain.

\begin{figure}\label{fig:CurrentCartoon}
	\includegraphics[width = 0.5 \textwidth]{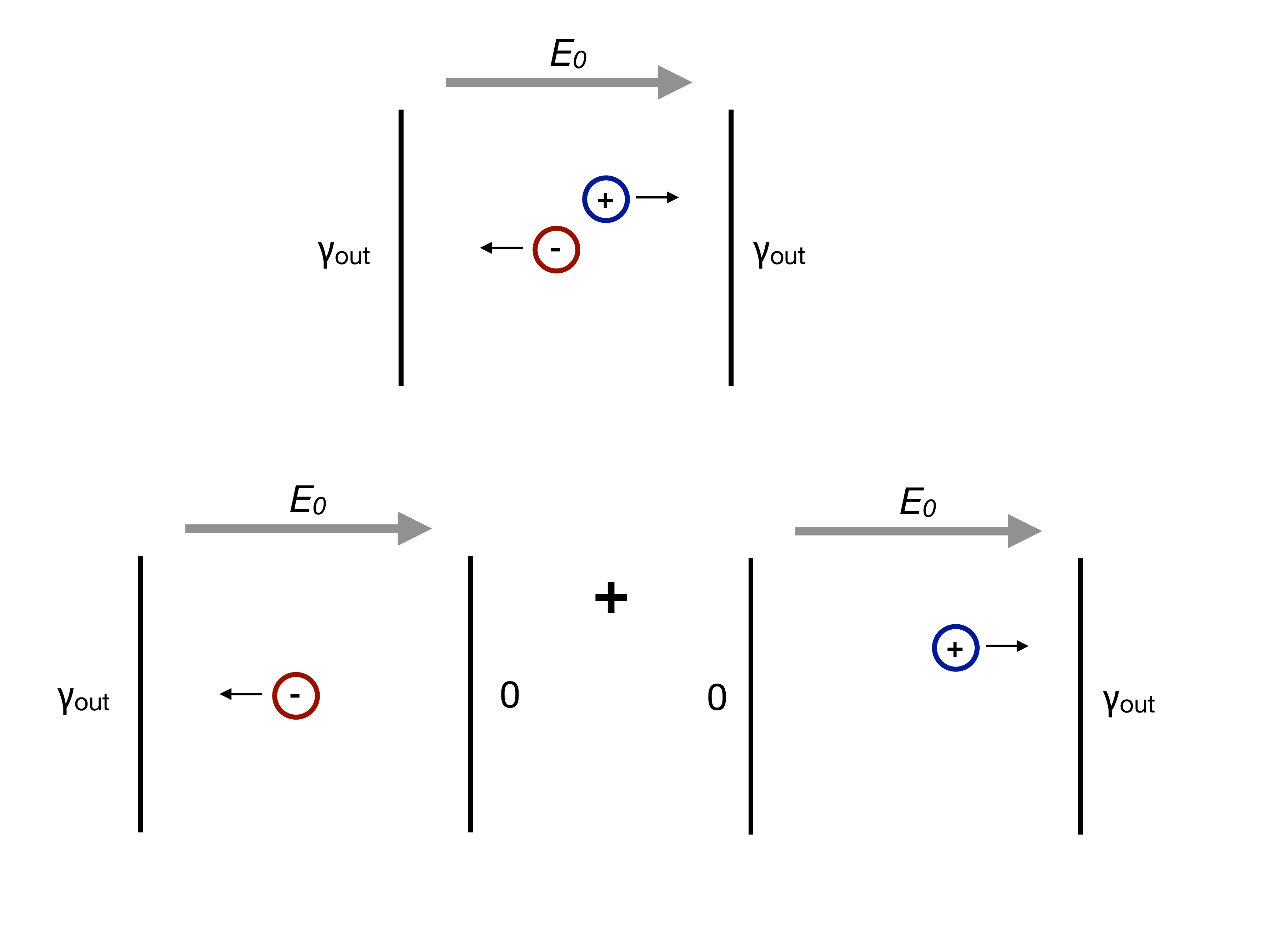}
	\caption{[Color online] Illustration of the decomposition of the system for electron-hole pairs. A realistic system consists of two collectors and describes the transport of electron-hole pairs. Since under the action of the electric field electrons and holes migrate to opposite collectors, the effective charge current can be modelled as the sum of the contributions from two chains with one collector. 
}
\end{figure}

\begin{figure}
	\includegraphics[width = 0.5 \textwidth]{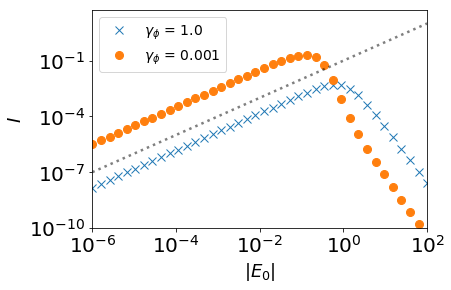}
	\caption{[Color online] Maximal charge current Eq.~\eqref{eq:Current} at the ST $\gamma_{out} = \gamma^{ST}$ as a function of the electric field gradient $E_0$ for a dephasing value above ($\gamma_\phi =1.0$) and below ($\gamma_\phi = 0.001$) the critical value $\tilde\gamma_\phi \sim 4\Omega/N$. For not too large electric field, the current is linear in the gradient (dotted line as visual guide). The chain has length $N=10$. The initial state is a Gaussian state Eq.~\eqref{eq:gauss3} centered in the middle of the chain, width one, and with zero momentum ($\Delta_0 = 1, n_0 = N/2, k_0 = 0$).
	}\label{fig:Current}
\end{figure}

\begin{figure}
	\includegraphics[width = 0.5 \textwidth]{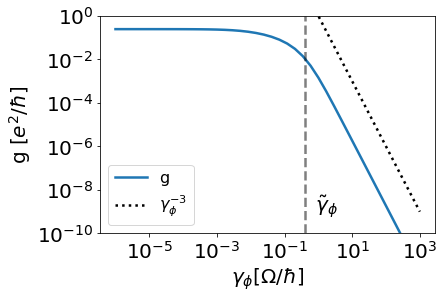}
	\caption{[Color online] Conductance at the superradiant transition $\gamma_{out} = \gamma^{ST}$ in function of the dephasing $\gamma_\phi$ for an electric field gradient below the critical value $E_0 < \tilde E_)$. Below the critical dephasing $\tilde\gamma_\phi \approx 4\Omega/N$ (dashed vertical line), the current is independent of the dephasing, and above the current is $\propto \gamma_\phi^{-3}$ (dotted line). The chain has length $N=10$.}\label{fig:Conductance}
\end{figure}

We can model the charge collection process by assuming that both ends of the chains are made of metallic contacts and thus act as collectors with rate $\gamma_{out}$. For electrons this amounts to assuming a Fermi-level larger than the system energy on the left (hard-wall), and lower on the right (sink). For holes, one should take the inverse system, with a collector on the left and hard-wall on the right. Therefore, assuming the complete symmetry between electrons and holes, the escape time for electrons should be exactly the same as for holes, but with reversed electric field gradient. The system with two collectors and an electron-hole pair can thus be decomposed into two chains, one with electrons and a collector on the right, and one with holes and a collector on the left, respectively, but with an electric field in the same direction in both cases as illustrated in Fig.~\ref{fig:CurrentCartoon}. We can hence define the charge current as
\begin{align}\label{eq:Current}
	I := e\left(\frac{1}{\tau(-E_0)} - \frac{1}{\tau(E_0)}\right).
\end{align}
We now compute the current $I$ for an initial excitation starting in the middle of the chain in order to account for the symmetry between holes and electrons, i.e., we consider an initial Gaussian state, see Eq.~\eqref{eq:gauss3}, with $n_0= (N+1)/2 = 5.5$ ($k_0 =0$ and $\sigma=1$ as previously). As shown in Fig.~\ref{fig:Current}, for $|E_0| \lesssim \tilde E_0$ the current is linear in the electric field gradient, in accordance with standard descriptions of current in solids \cite{Landauer1957,Gurvitz1996}. We can thus define a conductance as the proportionality constant $I = g V $ where $V =N |E_0| /e$ is the total potential drop. As is shown in Fig.~\ref{fig:Conductance}, the conductance $g\sim 0.25 e^2/\hbar$ is constant in the coherent regime ($\gamma_\phi < \tilde\gamma_\phi$ and $E_0 < \tilde E_0$), while it vanishes in the diffusive regime (according to our numerics we have found $g \sim \gamma_\phi^{-3}$ for large dephasing strength). 
It is important to note that in the coherent regime it is of same order of magnitude as the quantum of conductance in 1--d systems  $g_0 = 2e^2/h \sim 0.32 e^2/\hbar$ which characterizes the current in the Landauer formalism. 

We remark that the actual photo-current in TMOs must depend on the rate photo-excitation generation, i.e. on the incoming light. Our model does not include any excitation rate, and the current $I$ in Eq.~\eqref{eq:Current} must be understood as the maximal possible current in the single-excitation limit.

%


\section{Conclusions} 

It was shown that the superradiant enhancement of transport in one-dimensional chains is robust to the presence of moderate electric fields. This complements previous results that have shown superradiance to be robust against moderate levels of disorder \cite{Zhang2017b}, complex network geometries \cite{Zhang2017b}, and electron-electron interactions \cite{Kropf2019}. Hence, superradiance emerges as a very robust quantum coherent effect to maximize transport in small-scale systems. Moreover, we have shown that for dephasing and electric fields smaller than the mean level spacing, the transport is quantum coherent. The transport is optimal at a finite value of the electric potential because the particle acquires a momentum towards the sink, and thereby self-interference from reflection of the chain boundary are reduced. When the electric potential becomes too large, transport is suppressed due to the (partial) localization of the wavefunction.

The present results strongly impact on the possibility of exploiting coherence-enhanced transport mechanism to improve the photo-conversion efficiency of polar TMOs-based devices. As a paradigmatic case we start from the case of LaVO$_3$/SrVO$_3$ heterostructures, which have been recently suggested as potential candidate for quantum-coherent photo-conversion at ambient temperatures \cite{Kropf2019}. The electric potential along the transport axis was estimated to be around 0.08 eV{\AA}$^{-1}$ \cite{Assmann2013} and the distance between two sites is 7.849 {\AA} \cite{DeRaychaudhury2007}. Given an estimate of $\Omega \approx 200$meV \cite{DeRaychaudhury2007}, we obtain a value of $E_0\sim 3.13 \Omega \gg \tilde E_0 \sim 0.56 \Omega$. This would mean that we are deep in the localized regime, and quantum coherent transport is not possible. However, the overall electric potential can be controlled by the application of an external field. Our work suggests the possibility to enhance the current by applying a negative external bias to partially compensate the intrinsic potential slope and drive the system back to the quantum-coherent regime. 

More generally, the intrinsic electric field of polar TMO heterostructures, such as LaAlO3/SrTiO3, depends on the way the charges redistribute within the heterostructure itself. Typical values of electric fields can range between 0.01-0.1 eVA$^{-1}$ \cite{Song2018} thus offering the opportunity of engineering the local field in order to enhance the transport of photo-generated carriers. We must however emphasize that the existence, and also the magnitude of the optimal electric field is strongly dependent on the initial state configuration, the value of dephasing and of the coupling to the sink. All of these variables, including the magnitude of the intrinsic electric field are very difficult to estimate precisely and thus caution is required in the comparison with our theory.




\section*{Acknowledgements}
F.B. and C.M.K. acknowledge support by the Iniziativa Specifica INFN-DynSysMath. C.G. and F.B. acknowledge support from Universit\'a Cattolica del Sacro Cuore through D1, D.2.2 and D.3.1 grants. C.G. and F.B. acknowledge financial support from MIUR through the PRIN 2017 program (Prot. 20172H2SC4\_005). C.G. acknowledges financial support from MIUR through the PRIN 2015 program (Prot.2015C5SEJJ001). G.L.C. acknowledges the Conacyt project A1-S-22706.


\appendix

\section{Average transfer time expressed in function of the Liouvillian}\label{app:Liouvillian}
The integral in the average transfer time $\tau$, \eqref{eq:taudef}, can be expressed analytically in terms of the eigenvalues and eigenvectors of the associated Liouvillian $\mathcal{L}$, the superoperator defined as $\dot{\hat{\rho}} = \mathcal{L} \hat{\rho}$. Indeed, upon diagonalization of $\mathcal{L}$ yields
\begin{align*}
	\mathcal{L} = V D V^{-1} \;\; ; \;\; D = \sum_{n} E_n |n\}\{n|,
\end{align*}
with $\mathcal{L} |n\} = E_n |n\}$, the dynamics $\hat{\rho}(t) = e^{\mathcal{L}t}\hat{\rho}(0)$ read
\begin{align}\label{eq:Rho(t)_analytical}
	\hat{\rho}(t) = V \left( \sum_n e^{E_nt} |n\}\{n|\right)V^{-1}\hat{\rho}(0).
\end{align}
Inserting Eq.~\eqref{eq:Rho(t)_analytical} into the definition of $\tau$, \eqref{eq:taudef}, and noting that $\mathbb{R}(E_n) < 0 \;\; \forall n$ since we only have one unique decay channel, yields
\begin{align}\label{eq:tauana}
	\tau = \frac{\gamma_{out}}{\hbar \eta} \{N| V \left(\sum_n\frac{1}{E_n^2} |n\}\{n|\right)V^{-1}|N\},
\end{align}
Thus, the numerical evaluation of $\tau$ can be done by evaluating Eq.~\eqref{eq:tauana}.

\section{Average transfer time for localized sites}\label{app:Localized}

\begin{figure}
	\includegraphics[width = 0.5 \textwidth]{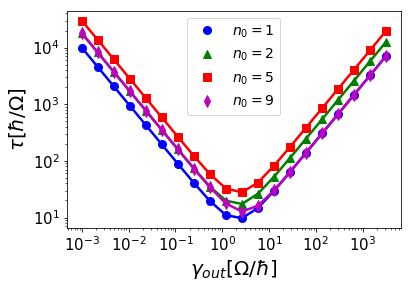}
	\caption{Average transfer time of an initially localized state $\ket{\psi_0} = \ket{n_0}$ in function of the coupling to the sink $\gamma_{out}$ for vanishing electric field and dephasing, $E_0 = 0,\gamma_\phi = 0$ for a chain of $N=10$ sites. The analytical expression Eq.~\eqref{eq:tau1} (full lines) perfectly agrees with the numerics (markers).}
	\label{fig:App1}
\end{figure}

For initial states localized on a single site $\ket{\psi_0} = \ket{n}$ we heuristically found an analytical description, Eq.~\eqref{eq:tau1}, of the average transfer time valid in the full quantum regime, i.e. for $\gamma_\phi = 0$ and $E_0=0$. The limit for $\gamma_{out} \ll \Omega$, Eq.~\eqref{eq:tauperturbn}, was derived using perturbation theory. In the opposite limit of large coupling to the sink we adapted the result obtained in \cite{Kropf2019} for $n=1$. As shown in Fig.~\ref{fig:App1} for $n=1,2,5,9$, the analytical expression Eq.~\eqref{eq:tau1} perfectly agrees with our numerical simulations. Interestingly, below the superradiant transition, $\gamma_{out} <\gamma^{ST}$, the fastest transport is always obtained starting from site $1$, which is the furthest from the sink. This quite counter-intuitive result arises from the quantum nature of the time-evolution. For any state not localized at the edge ($n_0 \neq 1$), part of the wave-function will initially move away from the sink due to ballistic spreading, which leads to a larger average transfer time. 

\begin{figure}
	\includegraphics[width = 0.5 \textwidth]{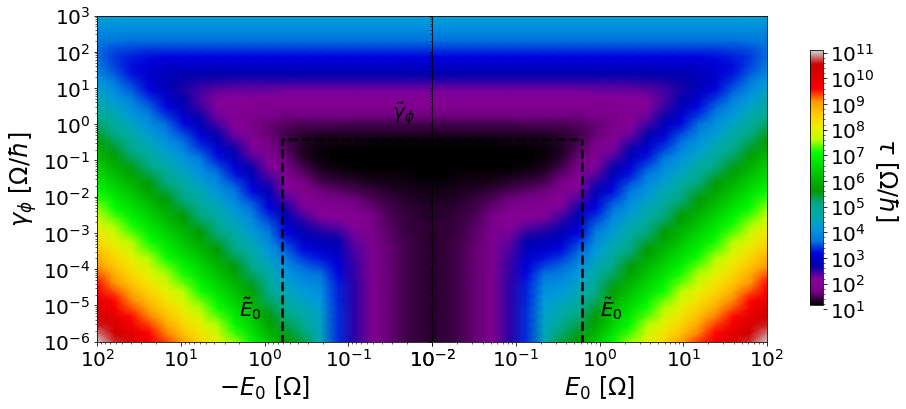}
	\caption{Average transfer time of an initially localized state $\psi_0 = \ket{n}$ in function of the electric field gradient $E_0$ and dephasing, $\gamma_\phi$ at the superradiant transition $\gamma_{out} = \gamma^{ST}$ for a chain of $N=10$ sites. A non-vanishing electric field is always detrimental for transport. There is an optimal value of the dephasing below $\tilde\gamma_\phi$, Eq.~\eqref{eq:dephcrit}. For reference to same Figure for the Gaussian initial state, Fig.~\ref{fig:Eopt2}, the critical electric field gradient $\tilde E_0$, Eq.~\eqref{eq:CriticalE0}, is indicated with vertical dahsed line.}
	\label{fig:App2}
\end{figure}

Furthermore, as we discussed in the main part of the manuscript, non-vanishing electric field are always neutral or detrimental to transport for a state initially localized on a single site $n$. As illustrated in Fig.~\ref{fig:App2} for the state $\ket{\psi_0} = \ket{3}$ and at the superradiant transition $\gamma_{out} = \gamma^{ST}$, for any value of the dephasing the optimal value of the electric field is $E_0 = 0$. In addition, there is a value of dephasing close to the transition to diffusive like-transport at $\tilde\gamma_{\phi}$ for which $\tau$ is minimized. This is consistent with our analysis for the Gaussian state. Changing the electric field allows to tune the overlap with the decaying states, or equivalently, can be used to give a momentum to a Gaussian state towards the sink. Both these mechanisms do not apply to localized states. At the same time, a non-vanishing dephasing leads to a general broadening of the energy levels and suppresses self-interferences. This leads to faster transport also for localized initial states for $\gamma_\phi$ below the threshold $\tilde\gamma_{\phi}$ to the diffusive-like regime. 

\vfill
 

\bibliographystyle{elsarticle-num}
\bibliography{Transport_Efield.bib}

\begin{thebibliography}{10}
\expandafter\ifx\csname url\endcsname\relax
  \def\url#1{\texttt{#1}}\fi
\expandafter\ifx\csname urlprefix\endcsname\relax\def\urlprefix{URL }\fi
\expandafter\ifx\csname href\endcsname\relax
  \def\href#1#2{#2} \def\path#1{#1}\fi

\bibitem{Chang2016}
S.~Chang, G.~D. Han, J.~G. Weis, H.~Park, O.~Hentz, Z.~Zhao, T.~M. Swager,
  S.~Grade{\v c}ak, Transition {{Metal}}-{{Oxide Free Perovskite Solar Cells
  Enabled}} by a {{New Organic Charge Transport Layer}}, ACS Appl. Mater.
  Interfaces 8~(13) (2016) 8511--8519.
\newblock \href {https://doi.org/10.1021/acsami.6b00635}
  {\path{doi:10.1021/acsami.6b00635}}.

\bibitem{Assmann2013}
E.~Assmann, P.~Blaha, R.~Laskowski, K.~Held, S.~Okamoto, G.~Sangiovanni, Oxide
  {{Heterostructures}} for {{Efficient Solar Cells}}, Phys. Rev. Lett. 110~(7)
  (2013) 078701.
\newblock \href {https://doi.org/10.1103/PhysRevLett.110.078701}
  {\path{doi:10.1103/PhysRevLett.110.078701}}.

\bibitem{Kropf2019}
C.~M. Kropf, A.~Valli, P.~Franceschini, G.~L. Celardo, M.~Capone, C.~Giannetti,
  F.~Borgonovi, Towards high-temperature coherence-enhanced transport in
  heterostructures of a few atomic layers, Phys. Rev. B 100~(3) (2019) 035126.
\newblock \href {https://doi.org/10.1103/PhysRevB.100.035126}
  {\path{doi:10.1103/PhysRevB.100.035126}}.

\bibitem{Ryndyk2016}
D.~A. Ryndyk, Landauer-{{B{\"u}ttiker Method}}, in: D.~Ryndyk (Ed.), Theory of
  {{Quantum Transport}} at {{Nanoscale}}: {{An Introduction}}, Springer
  {{Series}} in {{Solid}}-{{State Sciences}}, {Springer International
  Publishing}, {Cham}, 2016, pp. 17--54.
\newblock \href {https://doi.org/10.1007/978-3-319-24088-6_2}
  {\path{doi:10.1007/978-3-319-24088-6_2}}.

\bibitem{Landauer1957}
R.~Landauer, Spatial {{Variation}} of {{Currents}} and {{Fields Due}} to
  {{Localized Scatterers}} in {{Metallic Conduction}}, IBM J. Res. Dev. 1~(3)
  (1957) 223--231.
\newblock \href {https://doi.org/10.1147/rd.13.0223}
  {\path{doi:10.1147/rd.13.0223}}.

\bibitem{Pastawski1991}
H.~M. Pastawski, Classical and quantum transport from generalized
  {{Landauer}}-{{B{\"u}ttiker}} equations, Phys. Rev. B 44~(12) (1991)
  6329--6339.
\newblock \href {https://doi.org/10.1103/PhysRevB.44.6329}
  {\path{doi:10.1103/PhysRevB.44.6329}}.

\bibitem{Gurvitz1996}
S.~A. Gurvitz, Y.~S. Prager, Microscopic derivation of rate equations for
  quantum transport, Phys. Rev. B 53~(23) (1996) 15932--15943.
\newblock \href {https://doi.org/10.1103/PhysRevB.53.15932}
  {\path{doi:10.1103/PhysRevB.53.15932}}.

\bibitem{Zener1934}
C.~Zener, R.~H. Fowler, A theory of the electrical breakdown of solid
  dielectrics, Proceedings of the Royal Society of London. Series A, Containing
  Papers of a Mathematical and Physical Character 145~(855) (1934) 523--529.
\newblock \href {https://doi.org/10.1098/rspa.1934.0116}
  {\path{doi:10.1098/rspa.1934.0116}}.

\bibitem{Celardo2012}
G.~L. Celardo, F.~Borgonovi, M.~Merkli, V.~I. Tsifrinovich, G.~P. Berman,
  Superradiance {{Transition}} in {{Photosynthetic Light}}-{{Harvesting
  Complexes}}, J. Phys. Chem. C 116~(42) (2012) 22105--22111.
\newblock \href {https://doi.org/10.1021/jp302627w}
  {\path{doi:10.1021/jp302627w}}.

\bibitem{Zhang2017a}
Y.~Zhang, G.~L. Celardo, F.~Borgonovi, L.~Kaplan, Opening-assisted coherent
  transport in the semiclassical regime, Phys. Rev. E 95~(2) (2017) 022122.
\newblock \href {https://doi.org/10.1103/PhysRevE.95.022122}
  {\path{doi:10.1103/PhysRevE.95.022122}}.

\bibitem{Zhang2017b}
Y.~Zhang, G.~L. Celardo, F.~Borgonovi, L.~Kaplan, Optimal dephasing for
  ballistic energy transfer in disordered linear chains, Phys. Rev. E 96~(5)
  (2017) 052103.
\newblock \href {https://doi.org/10.1103/PhysRevE.96.052103}
  {\path{doi:10.1103/PhysRevE.96.052103}}.

\bibitem{Chavez2019}
N.~C. Ch{\'a}vez, F.~Mattiotti, J.~A. {M{\'e}ndez-Berm{\'u}dez}, F.~Borgonovi,
  G.~L. Celardo, Real and imaginary energy gaps: A comparison between single
  excitation {{Superradiance}} and {{Superconductivity}} and robustness to
  disorder, Eur. Phys. J. B 92~(7) (2019) 144.
\newblock \href {https://doi.org/10.1140/epjb/e2019-100016-3}
  {\path{doi:10.1140/epjb/e2019-100016-3}}.

\bibitem{Leegwater1996}
J.~A. Leegwater, Coherent versus {{Incoherent Energy Transfer}} and
  {{Trapping}} in {{Photosynthetic Antenna Complexes}}, J. Phys. Chem. 100~(34)
  (1996) 14403--14409.
\newblock \href {https://doi.org/10.1021/jp961448i}
  {\path{doi:10.1021/jp961448i}}.

\bibitem{Giusteri2015}
G.~G. Giusteri, F.~Mattiotti, G.~L. Celardo, Non-{{Hermitian Hamiltonian}}
  approach to quantum transport in disordered networks with sinks: {{Validity}}
  and effectiveness, Phys. Rev. B 91~(9) (2015) 094301.
\newblock \href {https://doi.org/10.1103/PhysRevB.91.094301}
  {\path{doi:10.1103/PhysRevB.91.094301}}.

\bibitem{Haken1973}
H.~Haken, G.~Strobl, An exactly solvable model for coherent and incoherent
  exciton motion, Z. Physik 262~(2) (1973) 135--148.
\newblock \href {https://doi.org/10.1007/BF01399723}
  {\path{doi:10.1007/BF01399723}}.

\bibitem{Breuer2002}
H.~P. Breuer, F.~Petruccione, The {{Theory}} of {{Open Quantum Systems}},
  {Oxford University Press}, 2002.

\bibitem{Celardo2009}
G.~L. Celardo, L.~Kaplan, Superradiance transition in one-dimensional
  nanostructures: {{An}} effective non-{{Hermitian Hamiltonian}} formalism,
  Phys. Rev. B 79~(15) (2009) 155108.
\newblock \href {https://doi.org/10.1103/PhysRevB.79.155108}
  {\path{doi:10.1103/PhysRevB.79.155108}}.

\bibitem{Johansson2013}
J.~R. Johansson, P.~D. Nation, F.~Nori, {{QuTiP}} 2: {{A Python}} framework for
  the dynamics of open quantum systems, Computer Physics Communications 184~(4)
  (2013) 1234--1240.
\newblock \href {https://doi.org/10.1016/j.cpc.2012.11.019}
  {\path{doi:10.1016/j.cpc.2012.11.019}}.

\bibitem{Plenio2008}
M.~B. Plenio, S.~F. Huelga, Dephasing-assisted transport: Quantum networks and
  biomolecules, New J. Phys. 10~(11) (2008) 113019.
\newblock \href {https://doi.org/10.1088/1367-2630/10/11/113019}
  {\path{doi:10.1088/1367-2630/10/11/113019}}.

\bibitem{Li2015}
Y.~Li, F.~Caruso, E.~Gauger, S.~C. Benjamin, `{{Momentum}} rejuvenation'
  underlies the phenomenon of noise-assisted quantum energy flow, New J. Phys.
  17~(1) (2015) 013057.
\newblock \href {https://doi.org/10.1088/1367-2630/17/1/013057}
  {\path{doi:10.1088/1367-2630/17/1/013057}}.

\bibitem{Izrailev1998}
F.~M. Izrailev, S.~Ruffo, L.~Tessieri, Classical representation of the
  one-dimensional {{Anderson}} model, J. Phys. A: Math. Gen. 31~(23) (1998)
  5263--5270.
\newblock \href {https://doi.org/10.1088/0305-4470/31/23/008}
  {\path{doi:10.1088/0305-4470/31/23/008}}.

\bibitem{Foerster1948}
T.~F{\"o}rster, {Zwischenmolekulare Energiewanderung und Fluoreszenz}, Ann.
  Phys. 437~(1-2) (1948) 55--75.
\newblock \href {https://doi.org/10.1002/andp.19484370105}
  {\path{doi:10.1002/andp.19484370105}}.

\bibitem{DeRaychaudhury2007}
M.~De~Raychaudhury, E.~Pavarini, O.~K. Andersen, Orbital {{Fluctuations}} in
  the {{Different Phases}} of {{LaVO}}\_\{3\} and {{YVO}}\_\{3\}, Phys. Rev.
  Lett. 99~(12) (2007) 126402.
\newblock \href {https://doi.org/10.1103/PhysRevLett.99.126402}
  {\path{doi:10.1103/PhysRevLett.99.126402}}.

\bibitem{Song2018}
K.~Song, S.~Ryu, H.~Lee, T.~R. Paudel, C.~T. Koch, B.~Park, J.~K. Lee, S.-Y.
  Choi, Y.-M. Kim, J.~C. Kim, H.~Y. Jeong, M.~S. Rzchowski, E.~Y. Tsymbal,
  C.-B. Eom, S.~H. Oh, Direct imaging of the electron liquid at oxide
  interfaces, Nature Nanotech 13~(3) (2018) 198--203.
\newblock \href {https://doi.org/10.1038/s41565-017-0040-8}
  {\path{doi:10.1038/s41565-017-0040-8}}.

\end{thebibliography}

\end{document}